\documentclass[twocolumn]{article}  
\usepackage{balance}  

\usepackage{amsmath}
\usepackage{amssymb}
\usepackage{booktabs}
\usepackage{graphicx}
\usepackage{subcaption}

\begin{document}

\title{FASTER: Fusion AnalyticS for public Transport Event Response}  
\date{}


\author{Sebastien Blandin, Laura Wynter, Hasan Poonawala, Sean Laguna\\
IBM Research\\
Singapore\\
\texttt{\{sblandin, lwynter, hasanp, slaguna\}@sg.ibm.com} \\
Basile Dura\\
Ecole Polytechnique\\
France\\
\texttt{basile.dura@polytechnique.edu} \\
}

\maketitle

\begin{abstract}
Increasing urban concentration raises operational challenges that can benefit from integrated monitoring and decision support. Such complex systems need to leverage the full stack of analytical methods, from state estimation using multi-sensor fusion for situational awareness, to prediction and computation of optimal responses.  The FASTER platform that we describe in this work, deployed at nation scale and handling 1.5 billion public transport trips a year, offers such a full stack of techniques for this large-scale, real-time problem. FASTER provides fine-grained situational awareness and real-time decision support with the objective of improving the public transport commuter experience. The methods employed range from statistical machine learning to agent-based simulation and mixed-integer optimization. In this work we present an overview of the challenges and methods involved, with details of the commuter movement prediction module, as well as a discussion of open problems.
\end{abstract}


\begin{section}{Introduction}
Efficient movement of people  in increasingly dense cities is one of the key challenges towards sustainable growth of urban areas throughout the world. Enabling effective response to incidents and unforeseen events requires real-time monitoring of the public transport network level of service, which in turn hinges on fine-grained real-time information on passenger movements.

While real-time information on vehicle movements is at the heart of traditional control centers, high-quality quantitative information on passenger movements is usually lacking. Indeed, while ticketing data would represent the most natural source of such information, it does not generally provide destination information when a passenger enters the network, is often not available for processing in real-time, and in dense networks does not indicate which route is chosen. Cameras possess valuable information on passenger counts on platforms but due to computational constraints are seldom processed to provide quantitative measures. 

Hence estimation of the network state in terms of the current passenger movements requires the fusion of multiple real-time data sources. In a real-world setting however, numerous challenges arise. The different sensing sources available have heterogeneous coverage, latency, and error statistics. Methods used to leverage the multi-modal sources must therefore be robust to different noise levels and time scales. Because of the real-time nature of the problem, they must also be fast and scalable.

The FASTER solution is motivated by the  constraints arising when applying mainstream AI techniques to operational settings.
\paragraph{Robustness in practice}
An imperative of mission-critical applications is that a minimal level of service is required in all conditions.   Because public transport systems are event-based in nature, via the dynamics of train arrivals and departures, lack of data due to failure of the sensors or IT network is often indistinguishable from a fault of the underlying physical system being monitored.

\paragraph{Consistency across heterogeneous use-cases}
A city-scale cyber-physical system needs to support  heterogeneous use-cases, from monitoring of crowd levels in sections of train platforms to offline analysis of daily network level of service. This requires that the underlying solution integrates a trade-off between optimal estimation methods for specific use-cases (real-time, offline), and global coherence of the estimates.

\paragraph{Complexity versus optimality}
With a goal of model explainability and robustness, important for instance in critical situations such as incident response, it is important to control the complexity of the models used, for instance by combining simple linear models in a multi-modal fusion framework, whose output can in turn be used by parsimonious agent-based engines.

\paragraph{Contributions}
The main contributions of this work include:
\begin{itemize}
    \item design of an end-to-end solution using machine learning, agent-based simulation  and mixed-integer optimization,
    \item novel methodologies employed in specific analytic modules such as the passenger movement prediction  model,
    \item implementation and deployment of the solution at city-scale with constraints from real-time and offline settings.
\end{itemize}
In Section~\ref{sec:relWork} we present an overview of related work. Section~\ref{sec:architect} provides a description of the system architecture. In Section~\ref{sec:CHMM} we present and evaluate a framework for commuter movements prediction. In Section~\ref{sec:expVal} we outline the observed benefits of such an integrated approach. We conclude with open problems in Section~\ref{sec:opProb}.
\end{section}
\begin{section}{Related work}\label{sec:relWork}

In the context of road  networks,  modeling  vehicle movements has leveraged techniques from sequential estimation and automatic control as early as the 90's, with the seminal work of~\cite{papageorgiou1991alinea}.
More recently,  applications of data assimilation and distributed planning  have benefited from the preponderance of smartphones, used  as sensors and  instruments of feedback, via guidance and incentives~\cite{mmReport,isttt,hoogendoorn2014integrated}. Similar methods have subsequently been employed for  public transport networks~\cite{liu2019stochastic}. Further, the availability of unstructured data  has allowed adding semantics to pure spatio-temporal representation of dynamical patterns~\cite{pereira2015using,kling2012city}.

Mobile traces have been used to analyze and predict movement patterns of people~\cite{ma2014opportunities, jiang2016timegeo, poonawala2016singapore,shao2016slicing,candia2008uncovering,herrera2010evaluation,jiang2017activity,kaiser2013enabling,ho2016distributed,reades2007cellular} with applications ranging from real-time congestion monitoring to land-use planning, using  techniques such as non-linear filtering and topological graph analysis. However, the spatial resolution of GPS and cellular sensing is often a limiting factor in indoor settings. 

On the other hand, wifi sensing has fine-grained spatial resolution. \cite{paper-51} proposes a system to estimate the number of passengers in public transport vehicles. In~\cite{6583443}  users' locations at a mass event are tracked using probe and other  wifi requests. In~\cite{baoyang,le2017real}, the authors build a system to passively ``sniff'' wifi signals of office workers  with an online SVM model to predict their length of stay. Wifi sensing finds further application in the retail sector. The authors of~\cite{lv2017big} present  a solution to predict the next place that a user will visit based on a Hidden Markov Model (HMM) framework. 

In~\cite{jiang2017trajectorynet} the authors propose a Recurrent Neural Network approach to classify GPS trip traces by transportation mode. Deep generative models have been explored in~\cite{lin2017deep}. The authors of~\cite{6415713} use  probe requests to reveal  underlying social relationships. In~\cite{Barbera:2013:SCU:2504730.2504742} the authors build snapshots of users at a large scale event. These new opportunities to  efficiently manage cities through the use of connected technology have led to the  definition of ``urban computing''~\cite{zheng2014urban,konishi2016cityprophet,zhou2018early}. 

Agent-based models have  benefited from  developments in machine learning leading to hybrid models~\cite{zhang2018real}.  The problem of inferring train arrivals and hence delays of public transport services was addressed  by~\cite{HORN201567} where  regional train timetables are inferred using cell phone data by detecting bursts in number of cell phones. However, their method would not work well on a dense urban metro system. The problem of detecting events of commuters \textit{left behind} in a subway system is addressed in~\cite{isttt}. The authors rely on offline farecard data, and estimate the most likely model assuming known distributions of passengers walking times. 

Related to our  goal of modeling passenger movements is the inference of users' trip activities  in a public transportation system. \cite{han2016activity} proposes a semi-continuous hidden Markov model framework. Activities are clustered using a Gaussian mixture that depends on the start time and duration of the activity. Similarly,~\cite{yin2018generative} applies an HMM framework to  activity classification. 

While  agent-based simulation has historically been focused on infrequent planning exercises, more recent endeavors have proven that they are now practical for real-time applications~\cite{luke2004mason,erath2012large,ben63real,horni2016multi}. Progress in classic problems such as vehicle routing~\cite{bertsimas2018online} and the use of surrogate and hybrid models~\cite{osorio2013simulation,brailsford2018hybrid} have pushed the field forward, as well as system implementations such as the use of high-performance computing allowing reaching nation scale~\cite{osogami2012ibm}.
\end{section}

\begin{section}{System overview}\label{sec:architect}
The FASTER system is a city-scale  solution providing situational awareness and decision support to monitor and manage a large-scale public transport network, in particular in terms of improving the response during incidents and events.

\begin{subsection}{System context}

The system described in this work ingests several heterogeneous data sources with varying levels of latency in order to build a comprehensive and fine-grained view of the ground conditions, raise early warnings and alerts during unexpected events, and compute optimized response plans to public transport incidents. Data sources include  structured and unstructured data such as CCTV streams, ticketing information, wifi traces, system data on the train locations, and quantities derived from cellular devices.

One of the  ways in which users interact with the FASTER system is through the  \textit{key performance indicators} (KPI) that the system produces and transmits to the command centre. The KPIs produced include  real-time  quantities such as estimated station platform crowd, dwell time delays and long aggregate passenger wait times. The estimates are updated every time new data  becomes available, so that users  have access to the most accurate information despite latency of some data sources.  This set of KPIs constitutes the common representation model for all analysis.

Breaking with the usual situation in that planning users and command centre users have different, segregated tools and methodologies to analyze the public transport system, FASTER offers  these two classes of users access to the same system, so that all information used, estimated or observed, is consistent across the real-time analyses and the planning studies. Figure~\ref{fig:FASTERUI} showcases a view of the key performance indicators.
\begin{figure}[htb!]
    \centering
    \includegraphics[width=\columnwidth]{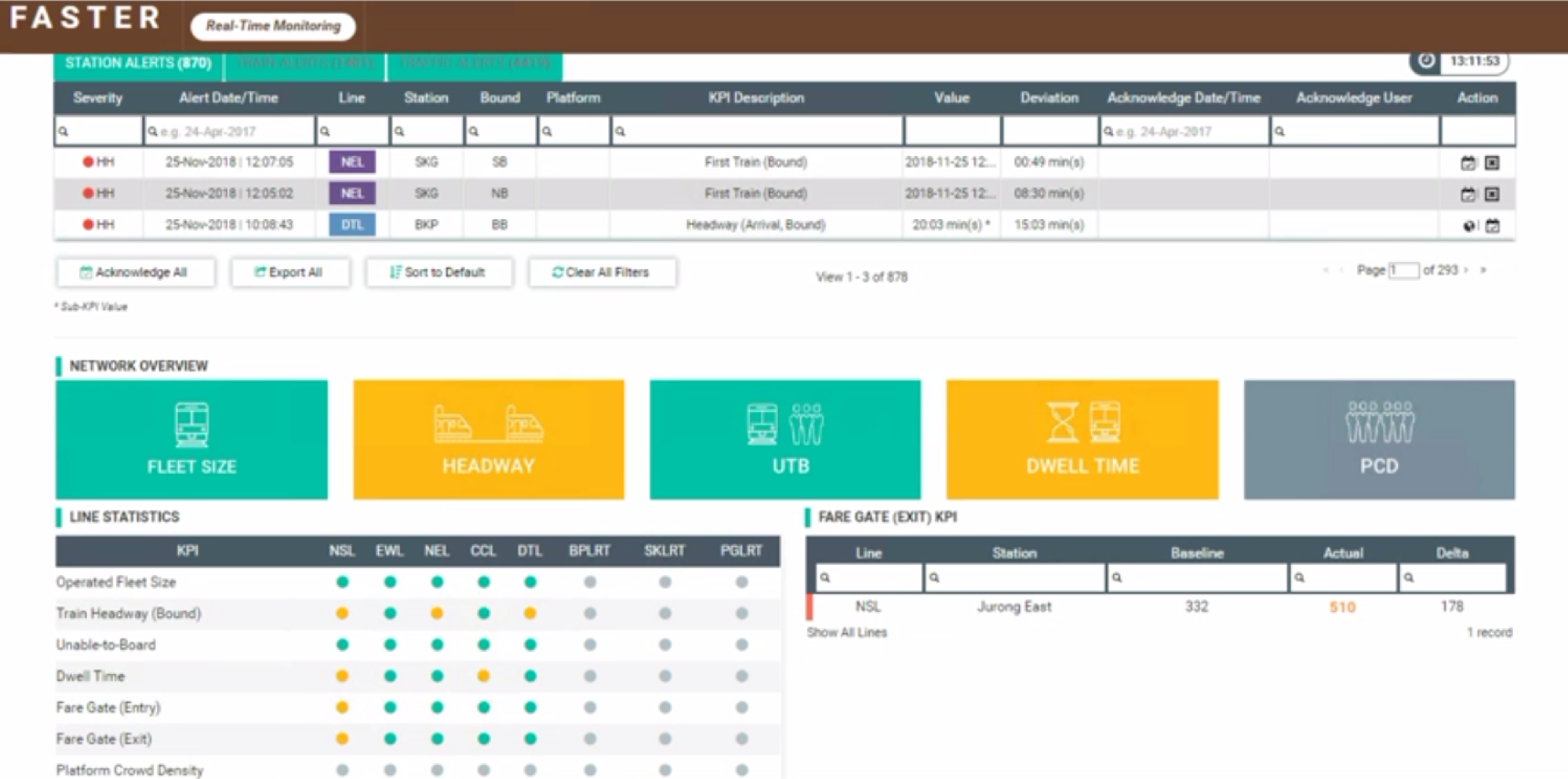}
    \caption{FASTER integrated real-time monitoring interface.}
    \label{fig:FASTERUI}
\end{figure}

Simulation and optimization functionalities are  integrated into the overall system using the  streaming data processing flow and the online KPI estimates. Thus the user can choose to analyze past actual days based on replaying the stored (real-time) estimates, simulate historical events with adjustments based also on the estimated values from the historical day, analyze ``typical day'' scenarios in the aggregate, or investigate more prospective hand-crafted scenarios designed from an arbitrary base-case. Users  can  run the estimated or simulated scenario forward in time, or they may trigger the mixed-integer optimization routine to find an improved solution  based on a given pool of resources and one or more pre-defined metrics. For instance, the optimization module can  recommend emergency bus routes  and schedules when an incident occurs.
\end{subsection}
\begin{subsection}{Lambda architecture}

The FASTER solution relies on a lambda architecture to ingest on the order of $1$ TB of data daily and serve all classes of users according to their requirements. Processed data feeds contribute to updating the common representation in the form of aggregate KPIs, which supports all of the applications such as prediction, alerts, production of optimal response plans, and playback analysis.

The  architecture, see Figure~\ref{fig:fasterRealTime}, includes both a batch layer and a speed layer with the speed layer focused on real-time monitoring and decision support, while the batch layer orchestrates heavy processing, simulation and calibration jobs.
\begin{figure}[htb!]
\centering
\includegraphics[width=\columnwidth]{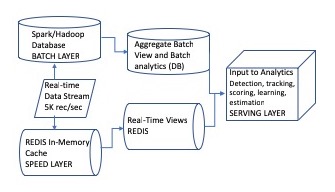}
\caption{Lambda architecture.}
\label{fig:fasterRealTime}
\end{figure}

The low-latency data feeds consumed include data from wifi-enabled devices, CCTV cameras, and train locations. 
\end{subsection}

\begin{subsection}{Reconciliation engine}
The FASTER solution  provides a full digital twin of locations, people, and vehicles using a common referential. Each of these object types is augmented with appropriate KPIs. The reconciliation engine thus alleviates the intrinsic limitations of individual sources, such as the limited coverage of cameras, the positioning noise of indoor traces, and the latency of ticketing information.

In order to support transparent fusion of data sources as they become available, we make use of a principled framework relying on data-stream specific fundamentals such as linear models and entity resolution methods, that we then combine in a common reconciliation engine integrating the most likely current common representation, as well as specific applications requirements.

For each KPI, or related group of KPIs, the reconciliation operates at the level of coarse agent metrics, using methods inspired by (prior) linear pooling from the combination of experts literature. Here each expert is a learner tasked to maximize the accuracy of certain commuter metrics, such as point-to-point travel-time, or crowd density. The estimates are then re-aligned on a common spatio-temporal grid, and re-weighted according to  estimates of the reliability of each learner for this data feed and  the  quantities produced in previous time steps.

This method allows improving estimate accuracy, and  has a computational advantage in terms of providing the estimates on the quantities of interest. Using a single combined indirect sensing mechanism facilitates the updates and reasoning as well as error analysis. The end result is that the FASTER system  produces accurate estimates of quantities such as train occupancy or platform crowding, which are traditionally not available  due to the lack of direct sensing mechanisms.
\end{subsection}

\begin{subsection}{Example: demand-supply gap estimation}
We illustrate the system design philosophy by describing below how a complex high-level KPI, the \textit{demand-supply gap} (DSG), is computed, based on estimates provided as part of the common referential in the form of low-level KPIs.

The demand-supply gap, expressed as the count of passengers unable to board a train at a given point in time, is a key metric of network level of service. However indirect methods such as network simulation only provide low-accuracy estimates, and no sensor provides a complete measurement of that quantity. In particular CCTV provides observations on portions of the platform and is notoriously difficult to use for measuring accurately the demand-supply gap. Ticketing data provides only the entry counts, reflecting the demand rather than the demand-supply gap.

In order to estimate the demand-supply gap, we rely on crowd level estimates, that are provided from linear models learning adaptively scaling parameters relating the number of connected device observations to actual crowd level in a supervised way. The DSG is then estimated using a discriminative classification method with the following feature set:
\begin{itemize}
\item count of commuters waiting to board a train,
\item count of commuters ``missing the train'', i.e. observed continuing to wait for the next train once a train departs,
\item waiting time third quartile and standard deviation,
\item train headway obtained by robust spectral clustering.
\end{itemize}
We highlight that we are estimating the macroscopic demand-supply gap, and not whether specific commuters will be left behind. We use greedy forward feature selection to select the most relevant features for model building. Since the datasets are highly skewed,  the vast majority of samples reflecting no DSG, we invoke a bootstrapping procedure to obtain an unbiased classification result.

Given the low number of DSG events, it is unrealistic to rely solely on station-specific models for accurate estimation. On the other hand, given the lack of stationarity of the underlying processes across stations and times, we cannot readily train models across the entire dataset. We thus normalize the features across the entire dataset, and build a hierarchy of models. The models are trained in a top-down fashion  from a network-wide model for all stations to line specific models, with distinct models for each  line on the network, and finally fine-grained models for each unique station on the network.
\end{subsection}

\begin{subsection}{Scalable agent-based modeling}

The FASTER simulation engine ingests the state estimated by the machine learning models used to create the KPI representation and forwards them in time to offer a fully data-driven simulation. 

Computation time being an imperative, the agent-based framework integrates microscopic models with a generic mesoscopic formulation in the form of a queuing network, which has been shown to provide a good trade-off between accuracy and efficiency. A dominant computation cost in traditional agent-based routing being the computation of shortest paths, we rely on low frequency shortest path updates combined with event-based re-computation to account for sudden global or local changes.

During simulation, several metrics can be produced depending on the user's interest. Unnecessary storage of simulation data is avoided by a  dynamically generated   data structure containing only the data necessary for the user-specified metric. This reduces the memory usage of the simulator by around $15\%$ and results in moderate but noticeable speedup.

A generic incident model supporting the class of faults encountered triggers the optimization engine. In the interest of scalability, we parameterize the set of candidate response lines for a given incident as illustrated in Figure~\ref{fig:FASTERresponselines}. The set is parameterized by the number of response lines, the allowed amount of overlap, the maximal lengths, and the number of distinct train lines they are required to cover. This achieves a good trade-off between quality of the response plans, responsiveness of the system, and practicality of the lines for actual real-time operations.
\begin{figure}[htb!]
    \centering
    \includegraphics[width=\columnwidth]{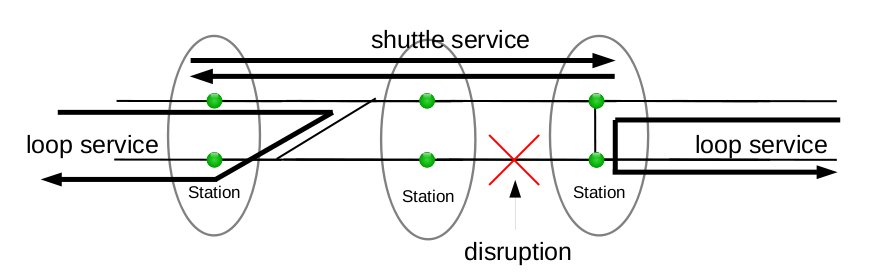} 
    \caption{Response lines; emergency train lines may ``loop'' around the incident or go through available tracks. Emergency bus lines are focused on re-connecting the network at a regional or connection to connection scale.}
    \label{fig:FASTERresponselines}
\end{figure}

In order to solve the optimal action plan problem, we first convert the discrete transport supply provided by individual train and bus services to a so-called continuous flow supply, and model the passenger movements as flows in a time expanded transportation network. The objective of the optimization routine is to produce emergency response bus routes and train schedules minimizing delay to impacted commuters.

We model the network using a classical time-expanded graph with a node $n_{u,t,l}$ representing a station $u$ at time $t$ for line $l$, and with arcs modeling service runtime, passenger walking time, and waiting time. For every group of $v_p$ passengers with same origin and start time, we add a commodity $p$ from a node $n_{o_p,s_p}$ (corresponding to origin station $o_p$ and start time $s_p$), to a destination node $z_{d_p}$ (corresponding to station $d_p$), with demand $v_p$. Taking the number of services $n_l$ on a line $l$ as a variable we get a mixed integer program solving the optimal action plan problem: 
\begin{eqnarray}
\min\sum_{p\in P,\, a\in A} f_{p,a}h_a \mbox{ s.t.}\label{f}\\
\sum_{n'\in N} f_{p,n'n} = \sum_{n'\in N} f_{p,nn'} \,\,\forall\,\, n \neq n_{o_p,s_p}, z_{d_p}, p\in P\label{fc1}\\
\sum_{n'\in N} f_{p,n'n} = v_p \mbox{ for } n=n_{o_p,s_p}\,\,\forall\,\, p\in P\label{fc2}\\
\sum_{n'\in N} f_{p,nn'} = -v_p \mbox{ for } n=z_{d_p}\,\,\forall\,\, p\in P\label{fc3}\\
\sum_{p\in P} f_{p,a} \leq c_ln_l/\tau_l  \,\,\,\forall\,\,\, \mbox{arc } a=\overrightarrow{n_{u,t,l}n_{v,t+r,l}}\in A.\label{cap}
\end{eqnarray}

\noindent with flow variables $f_{p,a}$, and where service capacities and headways are denoted $c_{l}$ and $\tau_{l}$, respectively, and with $a\in A$ the set of arcs. Equations \eqref{fc1}-\eqref{fc2}-\eqref{fc3} are the flow conservation constraints, and \eqref{cap} contains the service capacity constraints. The objective \eqref{f} is the total travel time of the passengers, various objective functions are supported in the system. Observe that a passenger uses a waiting arc $a=\overrightarrow{n_{v,t,l}n_{v,t+1,l}}$ only if line $l$ leaving from node $n_{v,t,l}$ is full. 

The mixed-integer program (MIP) is solved on multiple cores using CPLEX, and is followed by a local search step. If demand is well-met based on the result of the MIP optimization,  relative to the train and bus availability, the local search simply fine-tunes the result by either adjusting the rate of trains on existing lines or swapping out less-used lines for others. If the demand is not well-met or the number of available trains or buses is large, new train and bus lines are generated using a vehicle routing formulation,  based on the residual demand from the output of the  optimization,  to  replace or supplement lines in the current response plan.

A work queue is used to parallelize the simulation runs during the local search step. The simulation runs are assigned to threads in an asynchronously streaming manner such that minimal waiting is required by individual threads. The parallelism is designed such that each thread runs entirely independently from the other threads. These two features of the implementation allow for a very straightforward implementation of a distributed parallelism variation of the local search step, which generalizes to an API for running multiple simulations across hardware resources over the network.
\end{subsection}
\end{section}


\begin{section}{Modeling commuter dynamics}\label{sec:CHMM}

In this section we describe commuter movement models in a HMM framework, focusing on handling a continuous state-space, and scalability over large number of users.

\begin{subsection}{CHMM}

\begin{figure}[htb!]
    \centering
    \includegraphics[width=0.7\columnwidth]{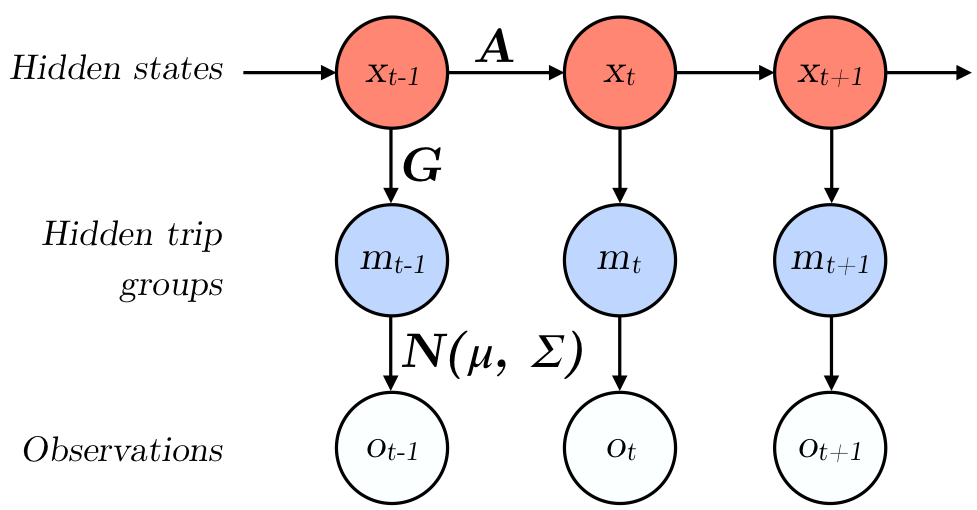} 
    \caption{Continuous HMM.}
    \label{fig:chmm}
\end{figure}
The \textit{continuous hidden Markov model} (CHMM)~\cite{han2016activity} is a HMM extension which considers clusters of hidden states, see Figure~\ref{fig:chmm}.
In our formulation, the index $t$ is the trip and the hidden variable $x_{t}$ is the exit station of the $t$ trip.
We denote $\mathbf{A}$ the transition matrix, so that $\mathbf{A} = \left\lbrace a_{i,j} \right\rbrace$ where $a_{i,j} = \mathbb{P}(x_{t+1} = j \: | \: x_t = i)$. Each state emits to a hidden cluster, according to a stochastic emission matrix $\mathbf{G} = \lbrace g_{i, k} \rbrace$ where $g_{i, k} = \mathbb{P}(m_t = k \: | \: x_t=i)$. Each cluster emits a continuous observation, according to a Gaussian distribution $o_t \sim \mathcal{N}(\mu_k, \Sigma_k)$. The CHMM model is thus fully parameterized by $\lambda = \left\lbrace \mathbf{A}, \mathbf{G}, \lbrace \mu_k, \Sigma_k \rbrace \right\rbrace$.

In our setting the model is trained on the observations given as a tuple including the time of entry in the network, duration of the  activity, i.e. the time out of the network, and the position of the exit station. We perform the parameter estimation via a variant of the Baum-Welch algorithm~\cite{dempster1977maximum}.
\end{subsection}

\begin{subsection}{Aggregate model}

The scalability of the CHMM approach can be substantially improved  by clustering similar users and building an aggregate model for each cluster. A naive approach and one that we use as a baseline is to compute a histogram representation of spatio-temporal frequency over a discretized domain for the relevant spatial and temporal features.
Specifically, in our baseline representation, the following spatio-temporal features are employed:
\begin{itemize}
    \item frequency of presence, over a discretized spatial domain,
    \item frequency of travel by time period, over a discretized temporal domain,
    \item spatial entropy, over a discretized spatial domain,
    \end{itemize}
 where the spatial entropy of user $u$ is defined as:
\begin{equation*}
    \textrm{entropy}(u) = - \sum_{s \in S_u} f_s \log f_s,
\end{equation*}
 $S_u$ is the set of stations visited by the user and $f_s$ is the frequency with which the user visited station $s$. The resulting adjacency matrix for a  subset of  $8000$ users using a euclidean distance metric is illustrated in Figure~\ref{fig:adjMatr}.
\begin{figure}[htb!]
\centering
\includegraphics[width=0.6\columnwidth]{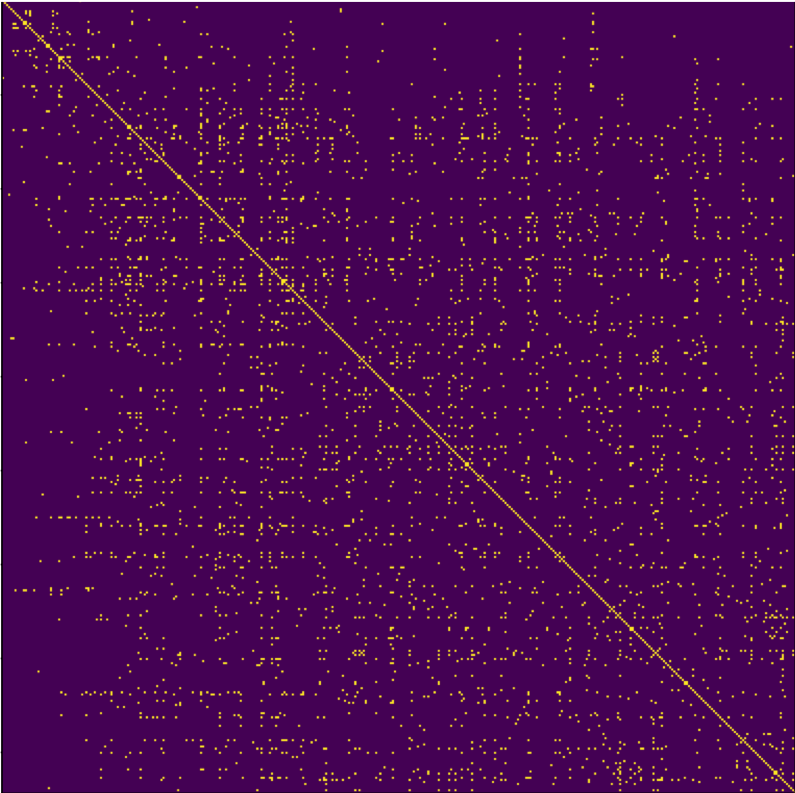}
\caption{Adjacency matrix for euclidean distance in spatio-temporal histogram-based feature space.}
\label{fig:adjMatr}
\end{figure}

The best available clustering of the adjacency matrix derived from the histogram distance, obtained using spectral clustering, is shown in Figure~\ref{fig:histclusters}. K-means and DBSCAN (not visualized here) produce significantly inferior clustering results on this dataset.
\begin{figure}[htb!]
\centering
\includegraphics[width=0.6\columnwidth]{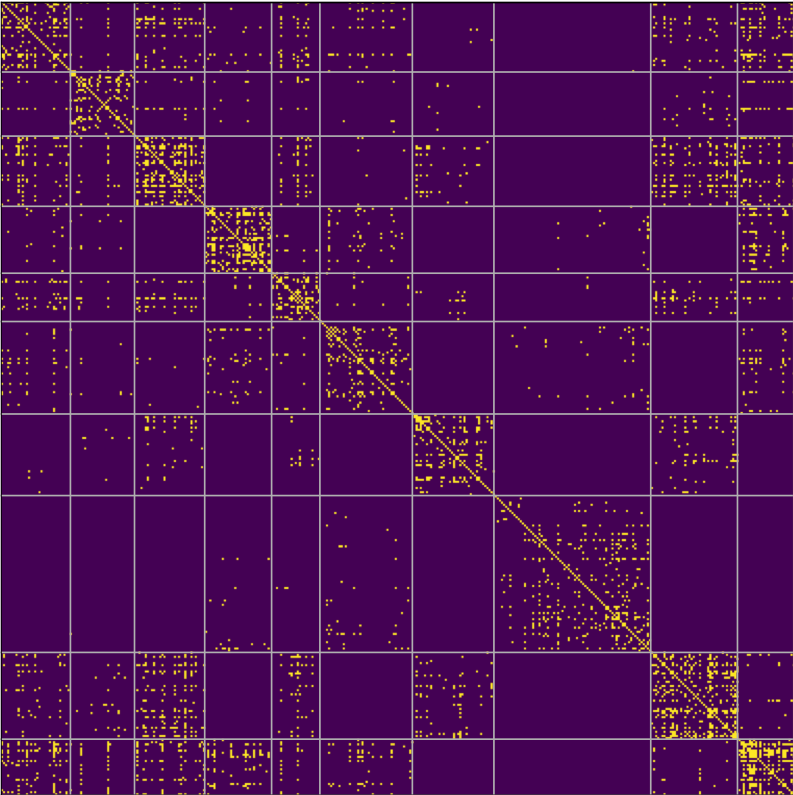}
\caption{Re-organized matrix after spectral clustering using euclidean distance on histogram-based feature space.}
\label{fig:histclusters}
\end{figure}

Note furthermore that the histogram representation of trips suffers from the drawback of not being able to distinguish trips with similar spatial patterns but differing temporal patterns; consider for example a user regularly making  the trip from A to B to C and one who travels from C to B to A. The histograms of such users will be identical in spite of vastly different temporal patterns.

We thus define a representation that describes the spatial and temporal patterns of users jointly. Specifically, we use the history of the user trips and the most likely CHMM corresponding to those trips. These CHMM are grouped into trip groups. Thus each user is represented by a Gaussian mixture.

Performing a new clustering on this CHMM-based representation of the users  requires a new distance metric. The appropriate metric in this case is the Kullback-Leibler (KL) divergence.
However, computing the KL divergence is computationally costly due to the lack of an analytical solution and as such, in the interest of scalability, we employ an approximation of the KL divergence in the form of the quadratic form distance. 

Defining the signature of a Gaussian mixture as:
\begin{equation*}
S^q = \lbrace \langle c_i^q, w_i^q \rangle, \: i=1..n \rbrace,
\end{equation*}
the Quadratic Form Distance (QFD) between two distributions reads $QFD(S^q, S^o) = \sqrt[]{(w_q-w_o) \cdot A_f \cdot (w_q-w_o)^\top}$, where $A_f$ is the \textit{similarity matrix} given by $a_{ij} = f(c_i, c_j)$, and $f$ is a pairwise distance such as $f_{-} (c_i, c_j) = -d(c_i, c_j)$ or $f_{g} (c_i, c_j) = e^{-\alpha d(c_i, c_j)^2}$.

Figure~\ref{fig:qfd-adjacency} shows the adjacency matrix resulting from spectral clustering performed on the pairwise gaussian quadratic form distance. Note that the clusters are far more homogeneous in size and with far fewer outliers than that obtained using spectral clustering on the histogram-based representation, shown in Figure~\ref{fig:histclusters}.
\begin{figure}[htb!]
    \centering
    \includegraphics[width=0.6\columnwidth]{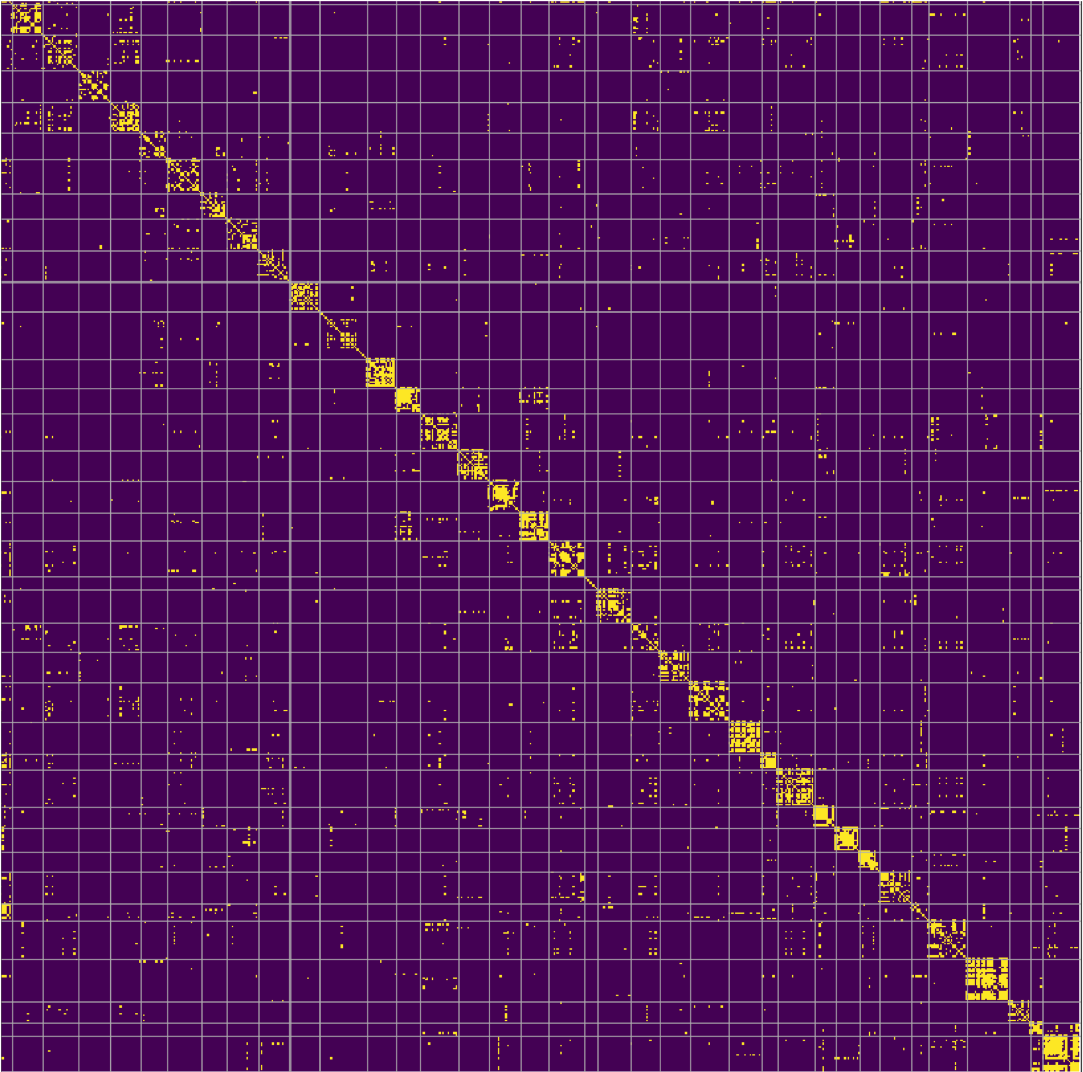}
    \caption{ Adjacency matrix obtained with spectral clustering using pairwise QFD.}
    \label{fig:qfd-adjacency}
\end{figure}
\end{subsection}
\end{section}


\subsection{CHMM model numerical results }\label{sec:numExp}

We consider a dataset of  $900$ million trips over a four months period, across $5$ train lines and $300$ bus routes.

The Gaussian mixture clustering of the activities in the CHMM is calibrated by maximizing the likelihood. On this dataset, when the number of Gaussian clusters  increases above ten, the log-likelihood drops substantially, due to overfitting of the model on the training set.
The hyper-parameters, namely the number of  hidden clusters, i.e. trip groups, optimized via a grid search, was set to  8.

\begin{figure}[htb!]
    \begin{subfigure}{\columnwidth}
        \centering
        \includegraphics[width=0.8\columnwidth]{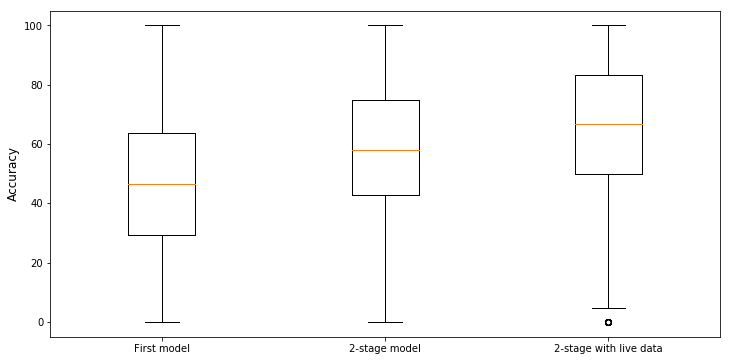}
        \caption{Full population}
    \end{subfigure}\\
    \begin{subfigure}{\columnwidth}
        \centering
        \includegraphics[width=0.8\columnwidth]{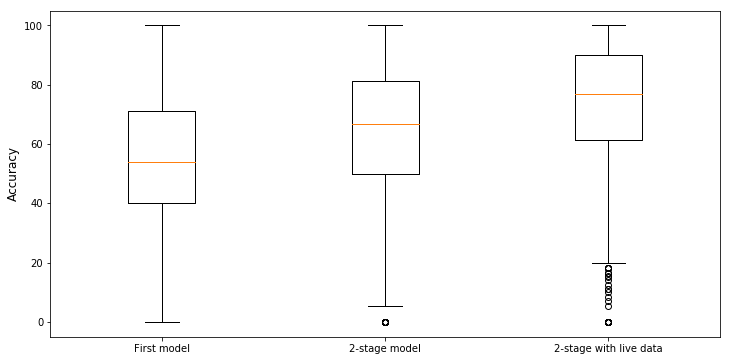} 
        \caption{$\textrm{Users having entropy} \leq 3$ (54.8\% of the population)}
    \end{subfigure}
    \caption{Accuracy of the proposed models, boxplots.}
    \label{fig:hmm-validation-accuracy}
\end{figure}

Figure~\ref{fig:hmm-validation-accuracy} shows the boxplots of the  model performance without clustering users; the box plots illustrate a  baseline CHMM in which only information from the previous trip $t$ is used for the $t+1$ prediction, the CHMM two stage-model, in which we augment the observations from the previous trip with the time of entry into the network and duration of the activity (in principle considered observations of the $t+1$ trip), and finally the two-stage model in its online version, in which the entry station, is also included as observation. A prediction is considered accurate if the station is within a range of $1$ km. The boxes show the median accuracy (middle red line), and the accuracy at the upper and lower quartiles (Q1=25\% and Q3=75\%). The inter-quartile range (IQR), defined as the accuracy range between Q3-Q1, is used to define the upper and lower horizontal lines as Q1-1.5(IQR) and Q3+1.5(IQR). Dots outside the horizontal lines represent the outliers.

A measurable improvement is observed when using the two-stage model compared to the CHMM baseline.


\begin{subsection}{Comparison of aggregate models performance}

Table~\ref{tab:KL-QFD} compares the prediction performance  of the 2-stage model (middle boxplot in 
Figure \ref{fig:hmm-validation-accuracy})
using the (symmetric) Kullback-Leibler and Quadratic Form Distances for the clustering.
\begin{table}[htb!]
    \centering
    \begin{tabular}{|l|c|c|}
        \hline
        & \textbf{KL} & \textbf{QFD}\\
        \hline
        \textbf{Mean}                   & 35.5\% & 33.2\% \\
        \textbf{Median}                 & 35.0\% & 30.0\% \\
        \hline
    \end{tabular}
    \caption{Comparison of cluster-level prediction accuracy for KL and QFD with spectral clustering, on 100 clusters.}
    \label{tab:KL-QFD}
\end{table}
As expected, the method using KL distance performs slightly better, but the loss of accuracy from the QFD approximation is very small, while the QFD approximation is obtained at a fraction of the computational cost.
\end{subsection}

\begin{section}{FASTER system  evaluation}\label{sec:expVal}
The FASTER solution described in this work has been implemented, deployed, and evaluated both in terms of  accuracy metrics for quantities of interest estimated by the system, as well as in terms of benefits obtained from the operational improvements enabled by the system. In this section we illustrate this validation process using an exemplary set of evaluations.

\begin{subsection}{Early warnings for real-time monitoring}

Real-time estimation of network conditions allows anomaly detection methods to raise alerts regarding situations that the control center should pay closer attention to. In Figure~\ref{fig:detection} we illustrate such a case of interruption of train services reported from 19:53pm to 20:23pm. The KPI illustrates clearly that the service was impacted  as early as 19:30pm, more than 20 minutes before the incident report was created.
\begin{figure}[htb!]
 \centering
 \includegraphics[width=0.7\columnwidth]{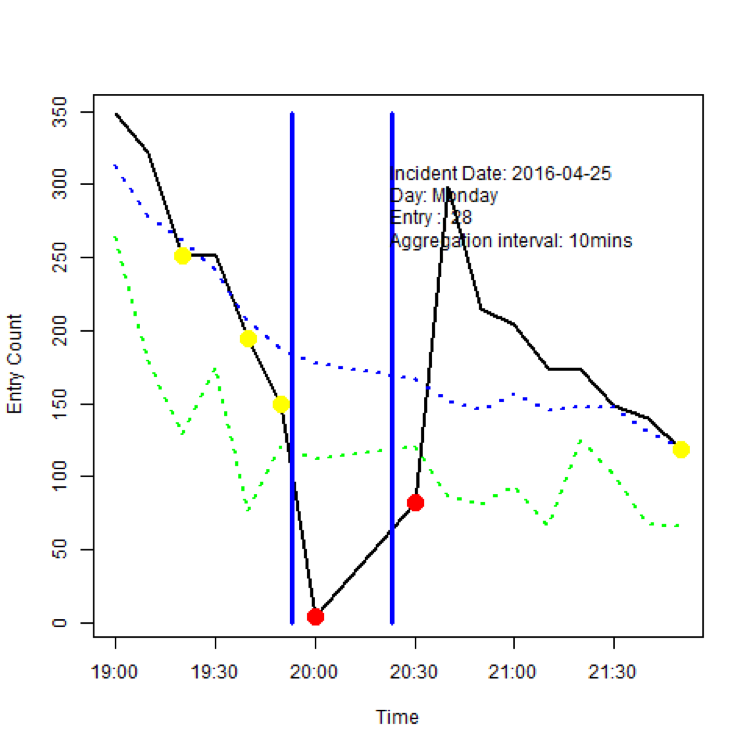}
 \caption{An anomalous trend in commuter crowding (solid black line) exceeding the medium (dotted blue line) and severe (dotted green line) alert levels is detected before manual incident reports (vertical blue lines).}
 \label{fig:detection}
\end{figure}

Such early warnings have been proven useful to mitigate the compounded impact of the incident as time progresses. We highlight that, as in most real-world implementations, the priority of anomaly detection methods is to maintain the number of false positive under a certain value, and maximize the number of true positive under this constraint.
\end{subsection}

\begin{subsection}{Daily estimation of demand-supply gap}

A key quantity in the monitoring of the quality of the transport service level is the demand-supply gap (DSG). The DSG measures the proportion of commuters intending to travel who are unable to board a train because it is full. Due to the difficulty of collecting fine-grained ground-truth  DSG estimates (i.e. how many trains commuters are forced to miss before boarding a train), we perform validation on the binary DSG detection problem (i.e. existence during a time period of a DSG event or not). 

We used  100K ground-truth DSG event labels (positive and negative instances) collected over a period of $8$ months at about $60$ stations, where a DSG event is declared if any passenger is forcedly left behind due to lack of capacity. In Table~\ref{table:modelHierarchy} we present Precision, Recall, and Accuracy, for detecting DSG for a family of models, running at the station, line, or network level.
\begin{table}[!htb]
	\centering
	\begin{tabular}{ccccc}
		\toprule
		Category & $\#$Models & Precision & Recall & Accuracy\\
		\midrule
		Network   & 1 & 75 & 72 & 98\\
		Line    & 1 to 10 & 77 & 72 & 98\\
		Station & 10 to 100 & 85 & 75 & 99\\
		\bottomrule
	\end{tabular}
	\caption{Performance of different model categories.}
	\label{table:modelHierarchy}
\end{table}

The statistical improvement in model accuracy obtained with more fine-grained models has to be balanced with the complexity associated with the maintenance of $100$ times more model instances and data streams. As these algorithms form the basis of an operational system, the importance of model maintenance is not to be neglected. We refer the interested reader to~\cite{Hasan2018} for more details on this model.
\end{subsection}
\begin{subsection}{Long-term analysis of level of service}
The system includes a long-term demand model which can be invoked in particular for the management of planned events. This parsimonious long-term predictive model, used for one-day ahead to one-year ahead, was shown to perform well across a number of special events, with less than $20\%$ error $90\%$ of the time. We illustrate here the performance on the case of a yearly event in Figure~\ref{fig:prediction}.
\begin{figure}[htb!]
    \begin{subfigure}{\columnwidth}
        \centering
        \includegraphics[width=\columnwidth]{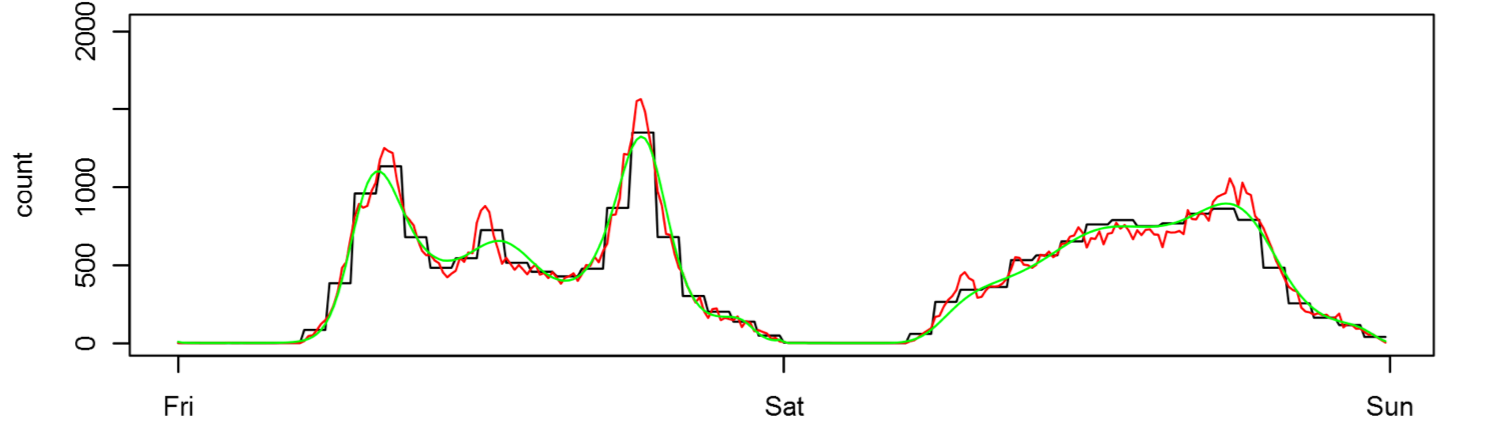}
        \caption{Offline prediction of crowding (green) at train station neighboring the event, at a $10$ min resolution, compared to actual situation (red), and typical day (black).}
    \end{subfigure}\\
     \begin{subfigure}{\columnwidth}
        \centering
        \includegraphics[width=\columnwidth]{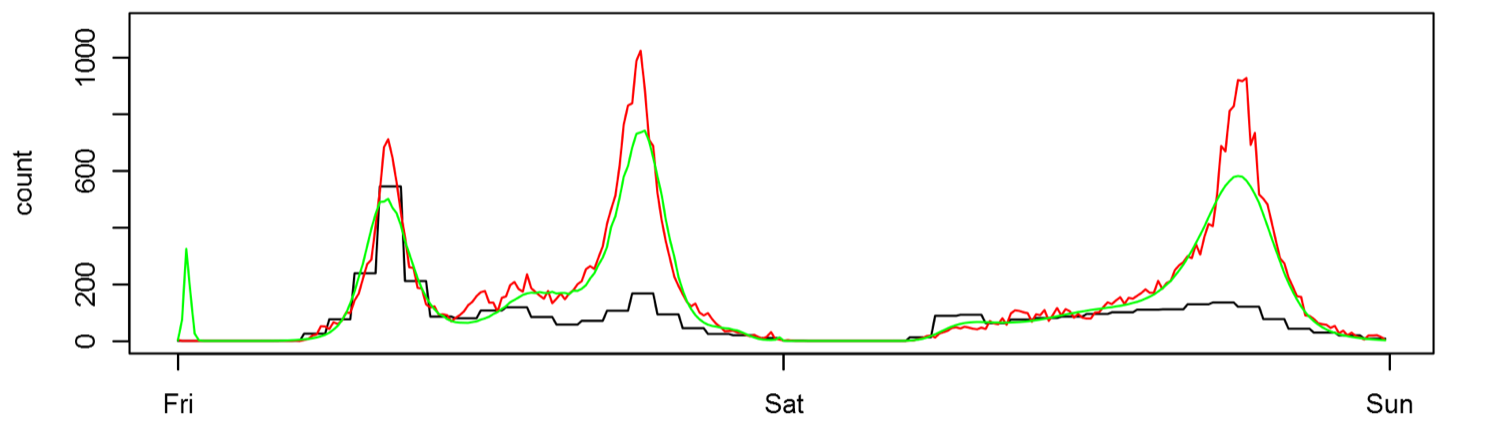}
        \caption{Offline prediction of crowding (green) at the train station closest to the event, at a $10$ min resolution, compared to actual situation (red), and typical day (black). Evening crowding is visible on the two days.}
    \end{subfigure}\\
    \caption{Performance of the predictive model, able to capture spatial distinctions in rare temporal variations due to recurrent events.}
    \label{fig:prediction}
\end{figure}

The model includes a hierarchy of calendar dependencies.  The first level includes weekday/weekend classes, the second level includes the  day of week, and the third level includes whether the day is a special event or not (National Day, New Year's Eve, etc.). Additive terms are then calibrated based on the full history, typically involving multiple years of data, to learn level-specific harmonics, considered as additive to the model from the previous level.
\end{subsection}

\begin{subsection}{Data-driven response plans}

Our agent-based implementation leads to a serial simulation of several millions of passengers and trains across $5$ lines at $5000$x speed sequentially. The system allows for meaningful explorations of alternatives during incidents and events based on the data-driven estimates of the current situation on the ground.
Table \ref{tab:errorSimu} gives the accuracy for a $6$ months period, in terms of the mean average error (MAE), the mean relative error (MRE) and the Bhattacharyya coefficients (BC) between the simulated and the smart card based travel-times of the passengers.

\begin{table}[htb!]
    \centering
    \begin{tabular}{ccccc}
    \toprule
         Passenger Set &  MAE (min.) & MRE (\%) & Avg. BC \\
         \midrule
         All & 4.9 & 18 & 0.93 \\
         Within one line & 2.5 & 19 & 0.95 \\
         \bottomrule
    \end{tabular}
    \caption{Agent-based simulation average travel time error. Simulated passengers travelling within a line are not subject to uncertainty at a transfers, hence the higher accuracy.}
    \label{tab:errorSimu}
\end{table}

During operations, the simulation - optimization engine evaluates on the order of $1000$ responses per incident, each being automatically generated based on the incident properties and available public transport resources. For simplicity,  properties of the demand, i.e. the  time-varying structure of the origin-destination matrix and the current station crowd densities, do not intervene in the \emph{a priori} design of the response plans, but only in their evaluation via simulation.
\begin{figure}[htb]
\centering
\includegraphics[width=0.8\columnwidth]{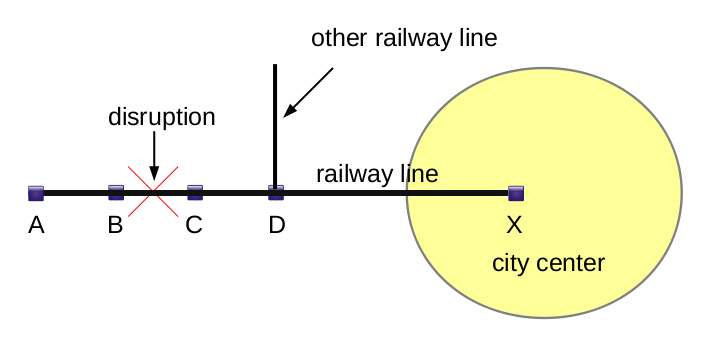}
\caption{Incident between station B and C in both directions, from 8am to 8:10am.}
\label{fig:incident2}
\end{figure}

For the incident illustrated in Figure~\ref{fig:incident2} occurring during the morning peak,  with the dominant direction of traffic being towards city center, the optimization engine proposes three diverse response plans with varying train headway, and standard emergency bus routes of varying headways, see Table~\ref{tab:accaggr}. In this example, given the heavy anisotropic demand and the localized spatio-temporal nature of the incident, the optimizer improves the situation by increasing the capacity towards city center. 

Given the heavy demand, the emergency bus lines, even operating at a very low headway of $1$ minute, are unable to accommodate the entire passenger flow. Hence even for short incidents ($10$ minutes here) it is important to manage the post-incident effects, such as by maintaining additional trains in service when the incident is over. Plan $3$ from Table~\ref{tab:accaggr} results in the best overall performance, reducing large delays as well as overcrowding. This is done by deploying two interleaving shuttle lines, one each from station C and D to and from the city center.
\begin{table}[htb!]
\centering
\begin{tabular}{|p{2.75cm}|p{1.2cm}|p{1.2cm}|p{1.2cm}|}
\hline
& Plan 1 & Plan 2 & Plan 3\\
\hline
Train service & C $\leftrightarrow$ X $\varphi=4$ & C $\leftrightarrow$ X $\varphi=4$ & C $\leftrightarrow$ X $\varphi=4$\\
& & & D $\leftrightarrow$ X $\varphi=8$ \\
\hline
\# additional trains& 0 & 0 & 10 \\
\hline
Shuttle bus service & B $\leftrightarrow$ C $\varphi=2$ & B $\leftrightarrow$ C $\varphi=1$ & B $\leftrightarrow$ C $\varphi=2$\\
\hline
Average delay (min) & 7 & 7 & 7 \\
\hline
Delay $\geq 20$ min (\#) & 300 & 280 & 200 \\
\hline
Overcrowding (min) & 20 & 19 & 10 \\
\hline
\end{tabular}
\caption{3 plans produced by the simulation optimization engine. Headway is denoted by the symbol $\varphi$.}
\label{tab:accaggr}
\end{table}

In other less constrained settings, the FASTER system has been able to produce non-standard response lines directly connecting the incident neighborhood with clusters of intended commuters destinations.  We refer the interested reader to~\cite{szabo2017data} for more details on the model.
\end{subsection}
\end{section}

\begin{section}{Conclusions and open problems}\label{sec:opProb}

In the context of the FASTER project, a number of important, yet often overlooked, challenges were encountered. Here we provide a succinct list of such problem statements, which, if addressed, will help facilitate more wide-spread adoption of agent-based techniques in large-scale operational systems.

\paragraph{Non-independent statistics}
Many sub-systems in  applications consume data produced by other sub-systems. This results in a deviation from traditional assumptions of statistical theory for the existence and convergence of estimators. A related  property of available data is that  the statistics of input data are  often non-stationary. Real-world applications would benefit from more principled research on such statistical challenges.

\paragraph{Real-time dominance}
One consequence of real-time independent sub-system interactions is that there is no opportunity to formally correct or update an estimate, since it is consumed as soon as it is produced. While multiple estimators differing by an allowed latency can be implemented, their significance decreases with the latency to the fastest estimator. Furthermore, high-latency estimators have to be either consistent with low-latency estimators, or in disagreement with transparent and sufficient evidence. In that context, efficient simulation - optimization of agent-based systems in real-time settings is of great value in the emerging area of digital twins and their use in operational control.

\paragraph{Scarcity of significant events}
A significant issue arising in large-scale real-world applications is that the situations of highest interest to users concern relatively rare circumstances. In contrast, while data-driven methods are obviously hindered by data scarcity, agent-based techniques have been considered relatively agnostic to the frequency of occurrence of the scenario considered. A significant gap remains between extreme data-driven methods performing very well in ideal conditions, and  principled methods with stable  average performance.
\end{section}
\section*{Acknowledgement}
  The authors would like to thank the Singapore Land Transport Authority (LTA) as well as  key colleagues from IBM who have worked on the FASTER system components: Vinod Bijlani, Panibhushan Shivaprasad and Paul Jose.
\bibliographystyle{plain}  
\balance  

\begin{thebibliography}{10}

\bibitem{Barbera:2013:SCU:2504730.2504742}
Marco~V. Barbera, Alessandro Epasto, Alessandro Mei, Vasile~C. Perta, and
  Julinda Stefa.
\newblock Signals from the crowd: Uncovering social relationships through
  smartphone probes.
\newblock In {\em Proceedings of the 2013 Conference on Internet Measurement
  Conference}, IMC '13, pages 265--276, New York, NY, USA, 2013. ACM.

\bibitem{mmReport}
A.~{\sc Bayen}, J.~{\sc Butler}, and A.~{\sc Patire}.
\newblock {Mobile Millennium final report}.
\newblock Technical report, 2011.

\bibitem{ben63real}
M.~{\sc Ben-Akiva}, M.~{\sc Bierlaire}, H.~{\sc Koutsopoulos}, and R.~{\sc
  Mishalani}.
\newblock Real-time simulation of traffic demand-supply interactions within
  {DynaMIT}.
\newblock {\em Transportation and network analysis: current trends: miscellanea
  in honor of Michael Florian}, 63:19--36, 2002.

\bibitem{bertsimas2018online}
Dimitris Bertsimas, Patrick Jaillet, and S{\'e}bastien Martin.
\newblock Online vehicle routing: The edge of optimization in large-scale
  applications.
\newblock 2018.

\bibitem{6583443}
B.~Bonn{\'{e}}, A.~Barzan, P.~Quax, and W.~Lamotte.
\newblock Wifipi: Involuntary tracking of visitors at mass events.
\newblock In {\em World of Wireless, Mobile and Multimedia Networks (WoWMoM),
  2013 IEEE 14\textsuperscript{th} International Symposium and Workshops on a},
  pages 1--6, June 2013.

\bibitem{brailsford2018hybrid}
Sally Brailsford, Tillal Eldabi, Martin Kunc, Navonil Mustafee, and Andres~F
  Osorio.
\newblock Hybrid simulation modelling in operational research: A
  state-of-the-art review.
\newblock {\em European Journal of Operational Research}, 2018.

\bibitem{candia2008uncovering}
Juli{\'a}n Candia, Marta~C Gonz{\'a}lez, Pu~Wang, Timothy Schoenharl, Greg
  Madey, and Albert-L{\'a}szl{\'o} Barab{\'a}si.
\newblock Uncovering individual and collective human dynamics from mobile phone
  records.
\newblock {\em Journal of physics A: mathematical and theoretical},
  41(22):224015, 2008.

\bibitem{6415713}
N.~Cheng, P.~Mohapatra, M.~Cunche, M.~A. Kaafar, R.~Boreli, and
  S.~Krishnamurthy.
\newblock Inferring user relationship from hidden information in wlans.
\newblock In {\em MILCOM 2012 - 2012 IEEE Military Communications Conference},
  pages 1--6, October 2012.

\bibitem{dempster1977maximum}
Arthur~P Dempster, Nan~M Laird, and Donald~B Rubin.
\newblock Maximum likelihood from incomplete data via the em algorithm.
\newblock {\em Journal of the royal statistical society. Series B
  (methodological)}, pages 1--38, 1977.

\bibitem{erath2012large}
Alex Erath, Pieter Fourie, Michael Van~Eggermond, Sergio Ord{\'o}{\~n}ez, Artem
  Chakirov, and Kay Axhausen.
\newblock Large-scale agent-based transport demand model for singapore.
\newblock In {\em 13th International Conference on Travel Behaviour Research
  (IATBR). Toronto: International Association for Travel Behaviour Research},
  2012.

\bibitem{han2016activity}
Gain Han and Keemin Sohn.
\newblock Activity imputation for trip-chains elicited from smart-card data
  using a continuous hidden markov model.
\newblock {\em Transportation Research Part B: Methodological}, 83:121--135,
  2016.

\bibitem{paper-51}
Marcus Handte, Muhammad~Umer Iqbal, Stephan Wagner, Wolfgang Apolinarski,
  Pedro~Jos{\'{e}} Marr{\'{o}}n, Eva Maria~Mu{\~{n}}oz Navarro, Santiago
  Martinez, Sara~Izquierdo Barthelemy, and Mario~Gonz{\'{a}}lez
  Fern{\'{a}}ndez.
\newblock {Crowd Density Estimation for Public Transport Vehicles}.
\newblock In {\em Workshop Proceedings of the EDBT/ICDT 2014 Joint Conference},
  Athens, Greece, 2014.

\bibitem{herrera2010evaluation}
Juan~C Herrera, Daniel~B Work, Ryan Herring, Xuegang~Jeff Ban, Quinn Jacobson,
  and Alexandre~M Bayen.
\newblock Evaluation of traffic data obtained via gps-enabled mobile phones:
  The mobile century field experiment.
\newblock {\em Transportation Research Part C: Emerging Technologies},
  18(4):568--583, 2010.

\bibitem{ho2016distributed}
Qirong Ho, Wenqing Lin, Eran Shaham, Shonali Krishnaswamy, Jingxuan Wang,
  Isabel~Choo Zhongyan, Amy She-Nash, et~al.
\newblock A distributed graph algorithm for discovering unique behavioral
  groups from large-scale telco data.
\newblock In {\em Proceedings of the 25th ACM International on Conference on
  Information and Knowledge Management}, pages 1353--1362. ACM, 2016.

\bibitem{hoogendoorn2014integrated}
Serge Hoogendoorn, Ramon Landman, Jaap van Kooten, Henk Taale, and Marco
  Schreuder.
\newblock Integrated network management amsterdam: Towards a field operational
  test.
\newblock In {\em Transportation Research Board 93\textsuperscript{rd} Annual
  Meeting}, number 14-2755, 2014.

\bibitem{HORN201567}
Christopher Horn and Roman Kern.
\newblock Deriving public transportation timetables with large-scale cell phone
  data.
\newblock {\em Procedia Computer Science}, 52:67--74, 2015.

\bibitem{horni2016multi}
Andreas Horni, Kai Nagel, and Kay~W Axhausen.
\newblock {\em The multi-agent transport simulation MATSim}.
\newblock Ubiquity Press London:, 2016.

\bibitem{jiang2017activity}
Shan Jiang, Joseph Ferreira, and Marta~C Gonz{\'a}lez.
\newblock Activity-based human mobility patterns inferred from mobile phone
  data: A case study of singapore.
\newblock {\em IEEE Transactions on Big Data}, 3(2):208--219, 2017.

\bibitem{jiang2016timegeo}
Shan Jiang, Yingxiang Yang, Siddharth Gupta, Daniele Veneziano, Shounak
  Athavale, and Marta~C Gonz{\'a}lez.
\newblock The timegeo modeling framework for urban motility without travel
  surveys.
\newblock {\em Proceedings of the National Academy of Sciences}, page
  201524261, 2016.

\bibitem{jiang2017trajectorynet}
Xiang Jiang, Erico~N de~Souza, Ahmad Pesaranghader, Baifan Hu, Daniel~L Silver,
  and Stan Matwin.
\newblock Trajectorynet: An embedded gps trajectory representation for
  point-based classification using recurrent neural networks.
\newblock {\em arXiv preprint arXiv:1705.02636}, 2017.

\bibitem{kaiser2013enabling}
Christian Kaiser and Alexei Pozdnoukhov.
\newblock Enabling real-time city sensing with kernel stream oracles and
  mapreduce.
\newblock {\em Pervasive and Mobile Computing}, 9(5):708--721, 2013.

\bibitem{kling2012city}
Felix Kling and Alexei Pozdnoukhov.
\newblock When a city tells a story: urban topic analysis.
\newblock In {\em Proceedings of the 20\textsuperscript{th} International
  Conference on Advances in Geographic Information Systems}, pages 482--485.
  ACM, 2012.

\bibitem{konishi2016cityprophet}
Tatsuya Konishi, Mikiya Maruyama, Kota Tsubouchi, and Masamichi Shimosaka.
\newblock Cityprophet: city-scale irregularity prediction using transit app
  logs.
\newblock In {\em Proceedings of the 2016 ACM International Joint Conference on
  Pervasive and Ubiquitous Computing}, pages 752--757. ACM, 2016.

\bibitem{le2017real}
Truc~Viet Le, Baoyang Song, and Laura Wynter.
\newblock Real-time prediction of length of stay using passive wi-fi sensing.
\newblock In {\em Communications (ICC), 2017 IEEE International Conference on},
  pages 1--6. IEEE, 2017.

\bibitem{lin2017deep}
Ziheng Lin, Mogeng Yin, Sidney Feygin, Madeleine Sheehan, Jean-Francois
  Paiement, and Alexei Pozdnoukhov.
\newblock Deep generative models of urban mobility.
\newblock {\em IEEE Transactions on Intelligent Transportation Systems}, 2017.

\bibitem{liu2019stochastic}
Yang Liu, Sebastien Blandin, and Samitha Samaranayake.
\newblock Stochastic on-time arrival problem in transit networks.
\newblock {\em Transportation Research Part B: Methodological}, 119:122--138,
  2019.

\bibitem{luke2004mason}
Sean Luke, Claudio Cioffi-Revilla, Liviu Panait, and Keith Sullivan.
\newblock Mason: A new multi-agent simulation toolkit.
\newblock In {\em Proceedings of the 2004 swarmfest workshop}, volume~8, pages
  316--327. Michigan, USA, 2004.

\bibitem{lv2017big}
Qiujian Lv, Yuanyuan Qiao, Nirwan Ansari, Jun Liu, and Jie Yang.
\newblock Big data driven hidden markov model based individual mobility
  prediction at points of interest.
\newblock {\em IEEE Transactions on Vehicular Technology}, 66(6):5204--5216,
  2017.

\bibitem{ma2014opportunities}
Huadong Ma, Dong Zhao, and Peiyan Yuan.
\newblock Opportunities in mobile crowd sensing.
\newblock {\em IEEE Communications Magazine}, 52(8):29--35, 2014.

\bibitem{osogami2012ibm}
T.~Osogami, T.~Imamichi, H.~Mizuta, T.~Morimura, R.~Raymond, T.~Suzumura,
  R.~Takahashi, and T.~Ide.
\newblock {IBM Mega traffic simulator}.
\newblock {\em IBM Res., Tokyo, Japan, IBM Res. Rep. RT0896}, 2012.

\bibitem{osorio2013simulation}
Carolina Osorio and Michel Bierlaire.
\newblock A simulation-based optimization framework for urban transportation
  problems.
\newblock {\em Operations Research}, 61(6):1333--1345, 2013.

\bibitem{papageorgiou1991alinea}
Markos Papageorgiou, Habib Hadj-Salem, and Jean-Marc Blosseville.
\newblock Alinea: A local feedback control law for on-ramp metering.
\newblock {\em Transportation Research Record}, 1320(1320):58--67, 1991.

\bibitem{pereira2015using}
Francisco~C Pereira, Filipe Rodrigues, and Moshe Ben-Akiva.
\newblock Using data from the web to predict public transport arrivals under
  special events scenarios.
\newblock {\em Journal of Intelligent Transportation Systems}, 19(3):273--288,
  2015.

\bibitem{poonawala2016singapore}
Hasan Poonawala, Vinay Kolar, Sebastien Blandin, Laura Wynter, and Sambit Sahu.
\newblock Singapore in motion: Insights on public transport service level
  through farecard and mobile data analytics.
\newblock In {\em Proceedings of the 22nd ACM SIGKDD International Conference
  on Knowledge Discovery and Data Mining}, pages 589--598. ACM, 2016.

\bibitem{reades2007cellular}
Jonathan Reades, Francesco Calabrese, Andres Sevtsuk, and Carlo Ratti.
\newblock Cellular census: Explorations in urban data collection.
\newblock {\em Pervasive Computing, IEEE}, 6(3):30--38, 2007.

\bibitem{shao2016slicing}
Jing Shao, Chen-Change Loy, Kai Kang, and Xiaogang Wang.
\newblock Slicing convolutional neural network for crowd video understanding.
\newblock In {\em Proceedings of the IEEE Conference on Computer Vision and
  Pattern Recognition}, pages 5620--5628, 2016.

\bibitem{Hasan2018}
Baoyang Song, Hasan Poonawala, Laura Wynter, and Sebastien Blandin.
\newblock Robust commuter movement inference from connected mobile devices.
\newblock In {\em Data Mining Workshop (ICDMW), 2018 IEEE International
  Conference on}. IEEE, 2018.

\bibitem{szabo2017data}
J{\'a}cint Szab{\'o}, Sebastien Blandin, and Charles Brett.
\newblock Data-driven simulation and optimization for incident response in
  urban railway networks.
\newblock In {\em Proceedings of the 16\textsuperscript{th} Conference on
  Autonomous Agents and MultiAgent Systems}, pages 819--827. International
  Foundation for Autonomous Agents and Multiagent Systems, 2017.

\bibitem{baoyang}
Viet~Le Truc, Baoyang Song, and Laura Wynter.
\newblock Ireal-time prediction of length of stay using passive wi-fi sensing.
\newblock In {\em Proceedings of the ICC}. IEEE, 2017.

\bibitem{yin2018generative}
Mogeng Yin, Madeleine Sheehan, Sidney Feygin, Jean-Fran{\c{c}}ois Paiement, and
  Alexei Pozdnoukhov.
\newblock A generative model of urban activities from cellular data.
\newblock {\em IEEE Transactions on Intelligent Transportation Systems},
  19(6):1682--1696, 2018.

\bibitem{zhang2018real}
Yan Zhang, Arnaud Grignard, Kevin Lyons, Alexander Aubuchon, and Kent Larson.
\newblock Real-time machine learning prediction of an agent-based model for
  urban decision-making.
\newblock In {\em Proceedings of the 17th International Conference on
  Autonomous Agents and MultiAgent Systems}, pages 2171--2173. International
  Foundation for Autonomous Agents and Multiagent Systems, 2018.

\bibitem{zheng2014urban}
Yu~Zheng, Licia Capra, Ouri Wolfson, and Hai Yang.
\newblock Urban computing: concepts, methodologies, and applications.
\newblock {\em ACM Transactions on Intelligent Systems and Technology (TIST)},
  5(3):38, 2014.

\bibitem{zhou2018early}
Jingbo Zhou, Hongbin Pei, and Haishan Wu.
\newblock Early warning of human crowds based on query data from baidu maps:
  Analysis based on shanghai stampede.
\newblock In {\em Big Data Support of Urban Planning and Management}, pages
  19--41. Springer, 2018.

\bibitem{isttt}
Yiwen Zhu, Haris~N. Koutsopoulos, and Nigel~H.M. Wilson.
\newblock Inferring left behind passengers in congested metro systems from
  automated data.
\newblock {\em Transportation Research Procedia}, 23:362--379, 2017.
\newblock Papers Selected for the 22\textsuperscript{nd} International
  Symposium on Transportation and Traffic Theory Chicago, Illinois, USA, 24-26
  July, 2017.

\end{thebibliography}

\end{document}